\documentstyle[12pt]{article}
\input tcilatex

\begin{document}

\title{The Radiation Transfer at a Layer of Magnetized Plasma With Random
Irregularities}
\author{N.A.Zabotin \and A.G.Bronin \and G.A.Zhbankov \\
\\
Rostov State University\\
194, Stachki Ave., Rostov-on-Don, 344090, Russia\\
E-mail: zabotin@iphys.rnd.runnet.ru\\
PACS: 52.35.Hr, 42.25.Fx, 94.20.Bb}

\begin{abstract}
The problem of radio wave reflection from an optically thick plane
monotonous layer of magnetized plasma is considered at present work. The
plasma electron density irregularities are described by spatial spectrum of
an arbitrary form. The small-angle scattering approximation \-in the
invariant ray coordinates is suggested for analytical investigation of the
radiation transfer equation. The approximated solution describing
spatial-and-angular distribution of radiation reflected from a plasma layer
has been obtained. The obtained solution has been investigated numerically
for the case of the ionospheric radio wave propagation. Two effects are the
consequence of multiple scattering: change of the reflected signal intensity
and anomalous refraction.
\end{abstract}

\maketitle

{\sf \newpage }

\section{Introduction}

Basic goal of the present work consists in derivation and analysis of the
transfer equation solution describing spatial-and-angular distribution of
radio radiation reflected from a plane stratified layer of magnetized plasma
with random irregularities.

The radiation transfer equation (RTE) in a randomly irregular magnetized
plasma was obtained in the work \cite{JETP92} under rather general initial
assumptions. In particular, the medium average properties were assumed
smoothly varying both in space and in time. In the work \cite{Zab93} the
radiation energy balance (REB) equation describing radiation transfer in a
plane stratified layer of stationary plasma with random irregularities has
been deduced. The invariant ray coordinates, allowing one to take into
account, by a natural way, refraction of waves and to represent the equation
in the most simple form, were used there. In the work \cite{Br93} it was
shown that the REB equation is a particular case of the radiation transfer
equation obtained in \cite{JETP92} and can be deduced from the latter by
means of transition to the invariant ray coordinates. REB equation, thus,
allows one to investigate influence of multiple scattering in a plane
stratified plasma layer on the characteristics of radiation. In particular,
it enables one to determine the spatial-and-angular distribution of
radiation leaving the layer if the source directivity diagram and
irregularity spatial spectrum are known. A few effects which require of wave
amplitudes coherent summation for their description (for example, phenomenon
of enhanced backscattering) are excluded from consideration. However, the
multiple scattering effects are much stronger, as a rule. This is true, in
particular, for the ionospheric radio propagation.

The numerical methods of the transfer equation solving developed in the
theory of neutron transfer and in the atmospheric optics appear useless for
the REB equation analysis. They are adapted, basically, to the solution of
one-dimensional problems with isotropic scattering and plane incident wave.
In a case of magnetized plasma the presence of regular refraction,
aspect-sensitive character of scattering on anisometric irregularities and
high dimension of the problem (the REB equation contains two angular and two
spatial coordinates as independent variables) complicate construction of the
effective numerical algorithm for its solving. In this situation it is
expedient to solve the REB equation in two stages. The first stage consists
of obtaining of the approximated analytical solution allowing one to carry
out the qualitative analysis of its properties and to reveal of its
peculiarities. At the second stage the numerical estimation methods can be
applied to the obtained analytical solution. This approach has been realized
at our work.

{}We begin the present paper from a detailed exposition of the invariant ray
coordinates concept. Then possibility to use of the small-angle scattering
in the invariant coordinates approximation is discussed. Two modifications
of the REB equation solution are obtained. The analysis of the obtained
solutions (both analitical and numerical) concludes the paper.

\section{Invariant ray coordinates and the radiation energy balance equation}

It is convenient to display graphically the electromagnetic wave propagation
in a plane-stratified plasma layer with the aid of the Poeverlein
construction \cite{Ginz67,Bud61}. We shall briefly describe it. Let the
Cartesian system of coordinates has axis $z$\ {} perpendicular and the plane 
$x0y$\ {} parallel to the plasma layer. We shall name such coordinate system
``vertical''. It is assumed that the vector of the external magnetic field \
{}$\vec H$ is situated in the plane $z0y$\ {}. Module of radius-vector of
any point inside of the unit sphere with centrum in the coordinate origin
corresponds to the value of refractive index\ $n_i(v,\alpha )${}, where $i=1$%
\ {} relates to the extraordinary wave, $i=$\ {} relates to the ordinary
one, $v=\omega _e^2/\omega ^2$\ {}, $\omega _e^2$\ {} is the plasma
frequency, \ {}$\omega ^2$ is the frequency of a wave, \ $\alpha ${} is the
angle between radius-vector and magnetic field $\vec H$\ {}. The refractive
index surface corresponding to a fixed value of \ \ $v${} and to all
possible directions of the radius-vector represents a rotation body about an
axis parallel to vector \ $\vec H$\ {} (see fig. 1).

Convenience of the described construction (in fact, this is an example of
coordinate system in space of wave vectors $\vec k${}) is become evident
when drawing the wave trajectory: it is represented by a straight line,
parallel to the axis $z$. This is a consequence of the generalized Snell
law, which also requires of equality of the fall angle and exit angle
onto/from a layer ($\theta $), and constantness of the wave vector azimuth
angle ($\varphi ${}). Note, that the crossing point of a wave trajectory
with the refractive index surface under given value of \ {}$v$ determines
current direction of the wave vector in a layer (it is anti-parallel to a
radius-vector) and current direction of the group speed vector (it coincides
with the normal to the refractive index surface). The projection of a wave
trajectory onto the plane $x0y$\ is a point which radius-vector has module $%
\sin \theta ${} and its angle with relation to axis \ $x$\ equals to $%
\varphi ${}. Thus, the coordinates $\theta $ and $\varphi ${} define
completely the whole ray trajectory shape in a plane layer and outside of it
and are, in this sense, invariant on this trajectory.

Radiation of an arbitrary point source of electromagnetic waves within the
solid angle $\theta \div \theta +d\theta ;\varphi \div \varphi +d\varphi $\
corresponds to the energy flow in the \ $\vec k${}-space inside of a
cylindrical ray tube parallel to axis \ $z${} with cross section $\sin
\theta d(\sin \theta )d\varphi =\sin \theta \cos \theta d\theta d\varphi $.
In case of regular (without random irregularities) plasma layer this energy
flow is conserved and completely determined by the source directivity
diagram: 
\begin{equation}
\label{eq1}P(z,\vec \rho ,\theta ,\varphi )=P_0(\vec \rho ,\theta ,\,\varphi
)\,\, 
\end{equation}
where $P$\ is energy flow density in the direction determined by angles $%
\theta ,\varphi $\ {} through the point $\vec \rho $\ {} on some base plane
situated outside of the layer parallel to it (in the ionosphere case it is
convenient to choose the Earth' surface as the base plane), \ $z$\ is
distance from the base plane (height in the ionosphere case). We shall
assume in the present paper that function $z(v)$\ {} is monotonous in the
region of wave propagation and reflection. If random irregularities are
absent and source of radiation is point, variable \ $\vec \rho $\ in (\ref
{eq1}) is superfluous, as the matter of fact, since unequivocal relation
between it and angles of arrival of a ray \ $\theta ,\varphi $\ {} exists.
When scattering is present the radiation energy redistributes over angular
variables \ {}$\theta ,\varphi $ and in space what is described by variable $%
\vec \rho ${}. The value of $P$\ {} satisfies in this case to the equation
of radiation energy balance \cite{Zab93,Br93}:{\footnotesize {\large {\rm \ }%
}} 
\begin{equation}
\label{eq2}
\begin{array}{l}
\frac d{dz}P(z,\theta ,\varphi ,\vec \rho )=\int \{-P(z,\vec \rho ,\theta
,\varphi )\sin \theta \cos \theta C^{-1}(z;\theta ,\varphi )\cdot \\ 
\cdot \sigma \left[ \alpha _0\left( \theta ,\varphi \right) ,\beta _0\left(
\theta ,\varphi \right) ;\alpha \left( \theta ^{\prime },\varphi ^{\prime
}\right) ,\beta \left( \theta ^{\prime },\varphi ^{\prime }\right) \right]
\sin \alpha (\theta ^{\prime },\varphi ^{\prime })\left| 
\frac{\partial (\alpha ,\beta )}{\partial (\theta ^{\prime },\varphi
^{\prime })}\right| + \\ +P\left[ z,\vec \rho -\vec \Phi (z;\theta ^{\prime
},\varphi ^{\prime };\theta ,\varphi ),\theta ^{\prime },\varphi ^{\prime
}\right] \sin \theta ^{\prime }\cos \theta ^{\prime }C^{-1}(z;\theta
^{\prime },\varphi ^{\prime })\cdot \\ 
\cdot \sigma \left[ \alpha _0\left( \theta ^{\prime },\varphi ^{\prime
}\right) ,\beta _0\left( \theta ^{\prime },\varphi ^{\prime }\right) ;\alpha
\left( \theta ,\varphi \right) ,\beta \left( \theta ,\varphi \right) \right]
\sin \alpha (\theta ,\varphi )\left| \frac{\partial (\alpha ,\beta )}{%
\partial (\theta ,\varphi )}\right| \}d\theta ^{\prime }d\varphi ^{\prime } 
\end{array}
\end{equation}
where $C(z;\theta ,\varphi )$\ is cosine of a ray trajectory inclination
angle corresponding to the invariant angles \ $\theta $\ and $\varphi
;\left| \partial (\alpha ,\beta )/\partial (\theta ,\varphi \right| $ is
Jacobean of transition from angular coordinates $\theta ,\varphi $\ \ to the
wave vector polar and azimuth angles $\alpha $\ \ and \ $\beta $\ in the
``magnetic'' coordinate system (which axis $0z$\ \ is parallel to the
magnetic field); \ 
$$
\sigma \left[ \alpha _0\left( \theta ,\varphi \right) ,\beta _0\left( \theta
,\varphi \right) ;\alpha \left( \theta ^{\prime },\varphi ^{\prime }\right)
,\beta \left( \theta ^{\prime }.\varphi ^{\prime }\right) \right] \equiv
\sigma \left[ \theta ,\varphi ;\theta ^{\prime },\varphi ^{\prime }\right] 
{\rm {\ }} 
$$
is scattering differential cross section describing intensity of the
scattered wave with wave vector coordinates $\alpha ,\beta $\ \ in magnetic
coordinate system (corresponding invariant coordinates are \ $\theta
^{\prime }$\ and \ $\varphi ^{\prime }$) which arises at interaction of the
wave with wave vector coordinates $\alpha _0,\beta _0$\ (invariant
coordinates $\theta ${\large {\rm {\ \ }}}and $\varphi $\ ) with
irregularities. Vector function $\vec \Phi (z;\theta ^{\prime },\varphi
^{\prime };\theta ,\varphi )$\ represents the displacement of the point of
arrival onto the base plane of a ray which has angular coordinates $\theta
^{\prime }$\ \ and $\varphi ^{\prime }$\ \ after scattering at level $z$\ \
with relation to the point of arrival of an incident ray with angular
coordinates $\theta ,\varphi $\ . It is essential that in a plane-stratified
medium the function \ $\vec \Phi $\ is determined only by smoothed layer
structure \ $v(z)$\ and does not depend on the scattering point horizontal
coordinate and also on coordinate $\vec \rho $\ \ of the incident and
scattered rays. Note also that ratio $\vec \Phi (z;\theta ,\varphi ;\theta
^{\prime },\varphi ^{\prime })=-\vec \Phi (z;\theta ^{\prime },\varphi
^{\prime };\theta ,\varphi )$\ \ takes place.

It is possible to check up that equation (2) satisfies to the energy
conservation law: when integrating over all possible for level $z$\ \ values
of $\theta ,\varphi $\ and $\,$all $\,\vec \rho $\ its right side turns into
zero. It is natural since in absence of true absorption the energy inside
the plasma layer does not collected.

\ Analyzing expression for the scattering differential cross section in a
magnetized plasma (see, for example, \cite{Akh74}), it is easy to be
convinced that the following symmetry ratio takes place: 
\begin{equation}
\label{eq3}\sigma \left[ \theta ,\varphi ;\theta ^{\prime },\varphi ^{\prime
}\right] \,n^2\cos \vartheta _g^{\prime }=\sigma \left[ \theta ^{\prime
},\varphi ^{\prime };\theta ,\varphi \right] n^{\prime 2}\cos \vartheta _g 
\end{equation}
where $\vartheta _g$\ {} is angle between the wave vector and group speed
vector, $n$\ {} is refractive index. Using (\ref{eq3}) the equation (\ref
{eq2}) can be presented as follows: 
\begin{equation}
\label{eq4}
\begin{array}{l}
{\rm \ }\frac d{dz}P(z,\vec \rho ,\theta ,\varphi )=\int Q(z;\theta ,\varphi
;\theta ^{\prime },\varphi ^{\prime }) \\ \left\{ P(z,\vec \rho -{\rm \ }%
\vec {\Phi (z};\theta ^{\prime },\varphi ^{\prime };\theta ,\varphi ),\theta
^{\prime },\varphi ^{\prime })-P(z,\vec \rho ,\theta ,\varphi )\right\}
d\theta ^{\prime }d\varphi ^{\prime } 
\end{array}
\end{equation}
where $Q(z;\theta ,\varphi ;\theta ^{\prime },\varphi ^{\prime })=\sigma
(\theta ,\varphi ;\theta ^{\prime },\varphi ^{\prime })C^{-1}(z,\theta
,\varphi )\sin \theta ^{\prime }\left| d\Omega _k^{\prime }/d\Omega ^{\prime
}\right| $\ , and quantity $\stackrel{\sim }{Q}(z;\theta ,\varphi ;\theta
^{\prime },\varphi ^{\prime })$\ $\equiv Q(z;\theta ,\varphi ;\theta
^{\prime },\varphi ^{\prime })\sin \theta \cos \theta ${} is symmetric with
relation to rearrangement of shaded and not shaded variables. The equation
REB in the form (\ref{eq4}) has the most compact and perfect appearance. It
is clear from physical reasons that (\ref{eq4}) has to have a unique
solution for given initial distribution $P_0(\vec \rho ,\theta ,\varphi ).$\
\ The obtained equation can be directly used for numerical calculation of
the signal strength spatial distribution in presence of scattering. However,
as it was noted at introduction already, this approach leads to essential
difficulties. Subsequent sections describe the method of construction of the
energy balance equation approximated analytical solution.

\section{Small-angle scattering approximation in the invariant ray
coordinates}

Let us consider the auxiliary equation of the following kind, which differs
from (4) only by absence of the dash over variable\ $\omega $\ marked by
arrow: 
\begin{equation}
\label{eq5}{\rm \ }\frac d{dz}P(z,\vec \rho ,\omega )=\int Q(z;\omega
;\omega ^{\prime })\left\{ P(z,\vec \rho +{\rm \ }\vec \Phi (z;\omega
;\omega ^{\prime }),\stackunder{\uparrow }{\omega })-P(z,\vec \rho ,\omega
)\right\} d\omega ^{\prime } 
\end{equation}
where designation $\omega =\left\{ \theta ,\varphi \right\} ,\,\,d\omega
=d\theta d\varphi ${} has been used for the sake of compactness. Equation (%
\ref{eq5}) can be easily solved analytically by means of Fourier
transformation over variable $\vec \rho $.\ \ The solution has the following
form: 
\begin{equation}
\label{eq6}\stackrel{\sim }{P}(z,\vec q,\omega )=P_0(\vec q,\omega
)S(z,0;\vec q,\omega ) 
\end{equation}
where $P_0(\vec q,\omega )${} is the Fourier image of the energy flow
density of radiation passing the layer in absence of scattering and the
value of $S$\ is defined by the expression 
\begin{equation}
\label{eq7}S(z_2,z_1,\vec q,\omega )=\exp \left\{ \stackrel{z_1}{\stackunder{%
z_1}{\dint }}dz^{\prime }\int d\omega ^{\prime }Q(z^{\prime };\omega ;\omega
^{\prime })\left[ e^{i\vec q\vec \Phi (z^{\prime };\omega ;\omega ^{\prime
})}-1\right] \right\} 
\end{equation}
One should note that integration over $z${} in this and subsequent formulae,
in fact, corresponds to integration along the ray trajectory with parameters 
$\theta ,\varphi $.\ The area of integration over $\omega ^{\prime }$\
includes rays which reflection level $h_r(\omega ^{\prime })>z$.

{}Let us transform now equation (\ref{eq4}) by the following way: 
\begin{equation}
\label{eq8}
\begin{array}{c}
\frac d{dz}P(z,\vec \rho ,\omega )=\int d\omega ^{\prime }Q(z;\omega ;\omega
^{\prime })\left\{ P(z,\vec \rho + 
{\rm \ }\vec \Phi (z;\omega ;\omega ^{\prime }),\omega )-P(z,\vec \rho
,\omega )\right\} + \\ +\int d\omega ^{\prime }Q(z;\omega ;\omega ^{\prime
})\left\{ P(z,\vec \rho +{\rm \ }\vec \Phi (z;\omega ;\omega ^{\prime
}),\omega ^{\prime })-\,P(z,\vec \rho +{\rm \ }\vec \Phi (z;\omega ;\omega
^{\prime }),\omega )\right\} 
\end{array}
\end{equation}
Its solution will be searched for in the form 
\begin{equation}
\label{eq9}P(z,\vec \rho ,\omega )=\stackrel{\sim }{P}(z,\vec \rho ,\omega
)+X(z,\vec \rho ,\omega ) 
\end{equation}
Thus, auxiliary equation (\ref{eq5}) allows to present the solution of the
equation (\ref{eq4}) in the form (\ref{eq9}). This is an exact
representation while some approximated expressions for quantities $\stackrel{%
\sim }{P}$\ \ and $X$ \ \ are not used.

{}By substituting of (\ref{eq9}) into the equation (\ref{eq4}) one can
obtain the following equation for the unknown function $X$\ : 
\begin{equation}
\label{eq10}
\begin{array}{l}
\frac d{dz}X(z,\vec \rho ,\omega )=\int d\omega ^{\prime }Q(z;\omega ;\omega
^{\prime })\{ 
\stackrel{\sim }{P}\left[ z,\vec \rho +{\rm \ }\vec \Phi (z;\omega ;\omega
^{\prime }),\omega ^{\prime }\right] -\, \\ - 
\stackrel{\sim }{P}\left[ z,\vec \rho +{\rm \ }\vec \Phi (z;\omega ;\omega
^{\prime }),\omega \right] \}+ \\ +\int d\omega ^{\prime }Q(z;\omega ;\omega
^{\prime })\{X\left[ z,\vec \rho +{\rm \ }\vec \Phi (z;\omega ;\omega
^{\prime }),\omega ^{\prime }\right] -X(z,\vec \rho ,\omega )\} 
\end{array}
\end{equation}
We shall assume now that the most probable distinction of angles $\omega
^{\prime }$\ \ and \ $\omega $\ is small. The heuristic basis for this
assumption is given by analysis of the Poeverlein construction (fig. 1). It
is easy to be convinced examining the Poeverlein construction that
scattering near the reflection level even for large angles in the wave
vector space entails small changes of the invariant angles $\theta ,\varphi $%
.\ \ This is especially true for irregularities strongly stretched along the
magnetic field (in this case the edges of scattered waves wave vectors form
circles shown in fig. 1 as patterns A and B). One should note also that the
changes of invariant angles $\theta ,\varphi $\ {} are certainly small if
scattering with small change of a wave vector direction takes place. This
situation is typical for irregularity spectra, in which irregularities with
scales more than sounding wave length dominate. Thus, the small-angle
scattering approximation in the invariant coordinates has wider
applicability area than common small-angle scattering approximation.

{}Scattering with small changes of $\theta ,\varphi $\ entails small value
of $\left| \vec \Phi \right| $.\ That follows directly both from sense of
this quantity and from the fact what $\left| \vec \Phi (z,\omega ,\omega
)\right| =0$. Let us make use of that to carry out expansion of quantity $X$%
\ {} at the right side of the equation (\ref{eq10}) into the Taylor series
with small quantities $\omega ^{\prime }-\omega $ \ {}and $\left| \vec \Phi
\right| $.{} Note that making similar expansion of function $P$\ \ at the
initial equation (\ref{eq4}) would be incorrect since function $P$ may not
to have property of continuity. For example, in case of a point source, $P_0$%
\ {} is a combination of $\delta ${}-functions. As it will be shown later,
the function $X$\ is expressed through $P_0$\ {}by means of repeated
integration and, hence, differentiability condition fulfils much easier for
it.

{}Leaving after expansion only small quantities of the first order, we
obtain the following equation in partial derivatives: 
\begin{equation}
\label{eq11}\frac \partial {\partial z}X(z,\vec \rho ,\omega )-A_\omega
(z,\omega )\frac \partial {\partial \omega }X(z,\vec \rho ,\omega )+A_{\vec
\rho }(z,\omega )\frac \partial {\partial \vec \rho }X(z,\vec \rho ,\omega
)=f({\rm \ }z,\vec \rho ,\omega ) 
\end{equation}
where%
$$
A_\omega (z,\omega )=\int d\omega ^{\prime }Q(z;\omega ;\omega ^{\prime
})(\omega ^{\prime }-\omega )\text{;} 
$$
$$
A_{\vec \rho }(z,\omega )=\int d\omega ^{\prime }Q(z;\omega ;\omega ^{\prime
})\vec \Phi (z,\omega ,\omega ^{\prime })\text{;} 
$$
$$
\begin{array}{l}
f( 
{\rm \ }z,\vec \rho ,\omega )=\int d\omega ^{\prime }Q(z;\omega ;\omega
^{\prime })\cdot \\ \cdot \left\{ \stackrel{\sim }{P}\left[ z,\vec \rho +%
{\rm \ }\vec \Phi (z;\omega ;\omega ^{\prime }),\omega ^{\prime }\right] -\,%
\stackrel{\sim }{P}\left[ z,\vec \rho +{\rm \ }\vec \Phi (z;\omega ;\omega
^{\prime }),\omega \right] \right\} 
\end{array}
$$
Here is the characteristic system for the equation (\ref{eq11}):%
$$
\frac{dX}{dz}=f({\rm \ }z,\vec \rho ,\omega )\text{;\thinspace \thinspace
\thinspace }\frac{d\vec \rho }{dz}=A_{\vec \rho }(z,\omega )\text{%
;\thinspace \thinspace \thinspace \thinspace }\frac{d\omega }{dz}=-A_\omega
(z,\omega ) 
$$
and initial conditions for it at $z=0$:\ 
$$
X=0\text{; }\,\vec \rho ^{\prime }=\vec \rho \text{;\thinspace }\omega
=\omega _0\text{.} 
$$
It is necessary to emphasize the distinction between quantity $\vec \rho
^{\prime }${}, which is a function of $z$, and invariant variable $\vec \rho 
$.

\ Solving the characteristic system we obtain%
$$
\,\omega =\omega (z,\omega _0)\text{,\thinspace \thinspace \thinspace
\thinspace \thinspace \thinspace \thinspace \thinspace \thinspace }\vec \rho
^{\prime }=\vec \rho -\stackrel{z_0}{\stackunder{z}{\dint }}dz^{\prime
}A_{\vec \rho }\left[ z^{\prime },\omega (z^{\prime },\omega _0)\right] \, 
$$
where $z_0$\ is $z$ coordinate of the base plane. It follows that 
\begin{equation}
\label{eq12}X(z_0,\vec \rho ,\omega )=\stackrel{z_0}{\stackunder{0}{\dint }}%
dz^{\prime }f\left\{ {\rm \ }z^{\prime },\vec \rho -\stackrel{z_0}{%
\stackunder{z^{\prime }}{\dint }}dz^{\prime \prime }A_{\vec \rho }\left[
z^{\prime \prime },\omega (z^{\prime \prime },\omega _0)\right] ,\omega
(z^{\prime },\omega _0)\right\} 
\end{equation}
Generally, expression (\ref{eq12}) gives the exact solution of the equation (%
\ref{eq11}). However, since we are already within the framework of the
invariant coordinate small-angle scattering approximation which assumes
small value of $A_\omega (z,\omega )${}, it is possible to simplify the
problem a little. Assuming $A_\omega \cong 0$\ and omitting index $0$ at
invariant coordinates $\omega ${}, we are caming to the following
approximate representation for function $X\,$: 
\begin{equation}
\label{eq13}
\begin{array}{l}
X(z_0,\vec \rho ,\omega )= 
\stackrel{z_0}{\stackunder{0}{\dint }}dz^{\prime }\{\stackrel{\sim }{P}%
\left[ z^{\prime },\vec \rho +{\rm \ }\vec \Phi (z^{\prime };\omega ;\omega
^{\prime })+\vec D(z_0,z^{\prime },\omega ),\omega ^{\prime }\right] - \\ 
\stackrel{\sim }{P}\left[ z^{\prime },\vec \rho +{\rm \ }\vec \Phi
(z^{\prime };\omega ;\omega ^{\prime })+\vec D(z_0,z^{\prime },\omega
),\omega \right] \} 
\end{array}
\end{equation}
\ where $\vec D(z_2,z_1,\omega )=\stackrel{z_2}{\stackunder{z_1}{\dint }}%
dz^{\prime }\int d\omega ^{\prime }Q(z^{\prime };\omega ;\omega ^{\prime
})\vec \Phi (z^{\prime },\omega ,\omega ^{\prime })$.

\ Thus, in the invariant coordinate small-angle scattering approximation the
solution of the REB equation (\ref{eq4}) is represented as a sum of two
terms (see (\ref{eq9})), the first of which is 
\begin{equation}
\label{eq14}
\begin{array}{l}
\stackrel{\sim }{P}\,(z_0,\vec \rho ,\omega )=\frac 1{\left( 2\pi \right)
^2}\,\int d^2q\ P_0(\vec q,\omega )\cdot \\ \cdot \exp \left\{ i\vec q\vec
\rho +\stackrel{z_0}{\stackunder{0}{\dint }}dz^{\prime }\int d\omega
^{\prime }Q(z^{\prime };\omega ;\omega ^{\prime })\left[ e^{i\vec q\vec \Phi
(z^{\prime };\omega ;\omega ^{\prime })}-1\right] \right\} 
\end{array}
\end{equation}
{}where \ $\frac 1{\left( 2\pi \right) ^2}\int d^2q\quad P_0(\vec q,\omega
)\exp (i\vec q\vec \rho )=P_0(\vec \rho ,\omega )${}, and the second one is
given by expression (\ref{eq13}).

\ The solution can be presented in the most simple form if one uses again
the smallness of quantity $\left| \vec \Phi \right| $\ and expands the
second exponent in the formula (\ref{eq14}) into a series. Leaving after
expansion only small quantities of the first order, one can obtain: 
\begin{equation}
\label{eq15}
\begin{array}{l}
P(z_0,\vec \rho ,\omega )\cong P 
\text{{\rm $_0\left[ \vec \rho +\vec D(z_0,0,\omega ),\omega \right] \,+$}}%
\stackrel{z_0}{\stackunder{0}{\dint }}dz^{\prime }\int d\omega ^{\prime
}Q(z^{\prime };\omega ;\omega ^{\prime })\cdot \\ \cdot \{ 
\stackrel{\sim }{P}\left[ z^{\prime },\vec \rho +{\rm \ }\vec \Phi
(z^{\prime };\omega ;\omega ^{\prime })+\vec D(z_0,z^{\prime },\omega )+\vec
D(z^{\prime },0,\omega ^{\prime }),\omega ^{\prime }\right] -\, \\ -%
\stackrel{\sim }{P}\left[ z^{\prime },\vec \rho +{\rm \ }\vec \Phi
(z^{\prime };\omega ;\omega ^{\prime })+\vec D(z_0,0,\omega ),\omega \right]
\}\text{ .} 
\end{array}
\end{equation}
The last operation is the more precise the faster value of {}$P_0(\vec
q,\omega )$ decreases under $\left| \vec q\right| \to \infty ${}. \ The
solution of the radiation energy balance equation obtained in the present
section in the form (\ref{eq9}), (\ref{eq14}), (\ref{eq13}), or in the form (%
\ref{eq15}), expresses the spatial-and-angular distribution of radiation
intensity passing layer of plasma with scattering through the
spatial-and-angular distribution of the incident radiation, that is, in
essence, through the source directivity diagram.

\section{Alternative approach in solving the REB equation}

{}The REB equation solving method stated in the previous section is based on
representation of quantity $P(z,\vec \rho ,\omega )$\ as a sum of the
singular part $\stackrel{\sim }{P}(z,\vec \rho ,\omega )$\ {} and the
regular one $X(z_0,\vec \rho ,\omega )${}. Regularity of the $X(z_0,\vec
\rho ,\omega )$\ {} has allowed one to use the expansion into the Taylor
series over variables $\vec \rho $\ {} and $\omega $\ at the equation (\ref
{eq10}) right side and to transform the integral-differential equation (\ref
{eq10}) into the first order partial derivative differential equation (\ref
{eq11}).

{}However, the stated approach is not the only possible. The REB equation
can be transformed right away using Fourier-representation of the function $%
P(z,\vec \rho ,\omega )${}:\ 
\begin{equation}
\label{eq16}P(z,\vec \rho ,\omega )=\frac 1{\left( 2\pi \right) ^2}\int
d^2qP(z,\vec q,\omega )\exp (i\vec q\vec \rho ) 
\end{equation}
Substitution of (\ref{eq16}) into (\ref{eq4}) gives the following equation
for quantity $P(z,\vec \rho ,\omega )$: 
\begin{equation}
\label{eq17}{\rm \ }\frac d{dz}P(z,\vec q,\omega )=\int d\omega ^{\prime
}Q(z;\omega ;\omega ^{\prime })\left\{ P(z,\vec q,\omega ^{\prime })\exp
\left[ i\vec q\vec \Phi (z;\omega ;\omega ^{\prime })\right] -P(z,\vec
q,\omega )\right\} 
\end{equation}
The quantity $P(z,\vec q,\omega )$\ is a differentiable function even when $%
P(z,\vec \rho ,\omega )$\ {} has peculiarities. Therefore, in the invariant
coordinate small-angle scattering approximation it is possible to use the
following expansion: 
\begin{equation}
\label{eq18}P(z,\vec q,\omega ^{\prime })\cong P(z,\vec q,\omega )+\frac{%
\partial P(z,\vec q,\omega )}{\partial \omega }(\omega ^{\prime }-\omega )\,%
\text{.} 
\end{equation}
Substituting (\ref{eq18}) in (\ref{eq17}) we obtain the partial derivative
differential equation 
\begin{equation}
\label{eq19}\frac \partial {\partial z}P(z,\vec q,\omega )-\stackrel{\sim }{A%
}(z,\vec q,\omega )\frac \partial {\partial \omega }P(z,\vec q,\omega
)-P(z,\vec q,\omega )\stackrel{\sim }{S}(z,\vec q,\omega )=0\, 
\end{equation}
where{} 
$$
\stackrel{\sim }{S}(z,\vec q,\omega )=\int d\omega ^{\prime }Q(z;\omega
;\omega ^{\prime })\left[ e^{i\vec q\vec \Phi (z;\omega ;\omega ^{\prime
})}-1\right] 
$$
$$
{\rm \ }\stackrel{\sim }{A}(z,\vec q,\omega )={\rm \ }\int d\omega ^{\prime
}Q(z;\omega ;\omega ^{\prime })\exp \left[ i\vec q\vec \Phi (z;\omega
;\omega ^{\prime })\right] (\omega ^{\prime }-\omega ){\rm \ }\text{.} 
$$
The characteristic system 
\begin{equation}
\label{eq20}\frac{d\omega }{dz}=-\stackrel{\sim }{A}(z,\vec q,\omega )\text{%
,\thinspace \thinspace \thinspace \thinspace \thinspace \thinspace }\,\frac{%
dP}{dz}=\stackrel{\sim }{S}(z,\vec q,\omega )P(z,\vec q,\omega ) 
\end{equation}
with initial conditions $P=P_0(\vec q,\omega )$, $\omega =\omega _0$\
{}at\thinspace $z=0$\ has the following solution: 
\begin{equation}
\label{eq21}P(z,\vec \rho ,\omega )=\frac 1{\left( 2\pi \right) ^2}\int
d^2qP_0(\vec q,\omega _0)\exp \left\{ i\vec q\vec \rho +\stackrel{z}{%
\stackunder{0}{\dint }}dz^{\prime }\stackrel{\sim }{S}\left[ z^{\prime
},\vec q,\omega (z^{\prime },\vec q,\omega _0\right] \right\} 
\end{equation}
This solution of the REB equation turns into the expression (\ref{eq14}) for 
$\stackrel{\sim }{P}$\ when $\stackrel{\sim }{A}(z,\vec q,\omega
)\longrightarrow 0$\ {}. But the latter limit transition corresponds to the
invariant coordinate small-angle scattering approximation used in the
previous section under derivation of (\ref{eq13}) and subsequent
expressions. Let us note, however, that in (\ref{eq21}), in contrast with (%
\ref{eq9}), any additional terms do not appear. It allows one to assume that
in used approximation the ratio 
\begin{equation}
\label{eq22}X(z,\vec \rho ,\omega )\ll P(z,\vec \rho ,\omega ) 
\end{equation}
is fulfiled. Additional arguments to the benefit of this assumption will be
presented in the following section.

\section{Analysis of the solution of the REB equation}

\ {}We shall show, first of all, that the obtained solution satisfies to the
energy conservation law. For this purpose it is necessary to carry out
integration of the left and right sides of (\ref{eq15}) over $\omega $\ \
and $\vec \rho $\ multiplied them previously by \ $\sin \theta \cos \theta $%
{}. The area of integration over angles is defined by the condition that
both wave \ $\omega $\ {} and wave \ $\omega ^{\prime }$\ {} achieve the
same level $z${} (since at level $z$\ {} their mutual scattering occurs). To
satisfy this condition one should add factors $\Theta \left[ h_r(\omega
)-z\right] $\ {}and $\Theta \left[ h_r(\omega ^{\prime })-z\right] $\ {}to
the integrand expression, where $\Theta (x)$\ {}is the Heviside step
function, $h_r(\omega )$\ \ {} is the maximum height which can be reached by
a ray with parameters $\theta ,\varphi ${}. Now integration can be expanded
over all possible values of angles, i.e., over interval \ {}$0\div \pi /2$\
for $\theta $\ {} and over interval $0\div 2\pi $\ {} for $\varphi ${}.
Then, (\ref{eq15}) becomes 
\begin{equation}
\label{eq23}
\begin{array}{l}
\int P(\omega )\sin \theta \cos \theta d\omega =\int P_0(\omega )\sin \theta
\cos \theta d\omega + \\ 
+\stackrel{z}{\stackunder{0}{\dint }}dz^{\prime }\int d\omega \int d\omega
^{\prime }\Theta \left[ h_r(\omega )-z^{\prime }\right] {}\Theta \left[
h_r(\omega ^{\prime })-z^{\prime }\right] \stackrel{\sim }{Q}(z^{\prime
};\omega ,\omega ^{\prime })\left[ P_0(\omega ^{\prime })-P_0(\omega
)\right] {\rm \ } 
\end{array}
\end{equation}
where \ $P(\omega ),P_0(\omega )$\ {} is a result of integration of $%
P(z_0,\vec \rho ,\omega )$\ {} and $P_0(\vec \rho ,\omega )$\ {}
correspondingly over variable $\vec \rho ${}.

{}Due to antisymmetry of the integrand expression with relation to
rearrangement of shaded and not shaded variables, the last term in (\ref
{eq23}) is equal to zero. Thus, equation (\ref{eq23}) reduces to 
\begin{equation}
\label{eq24}\int P(z_0,\vec \rho ,\omega )\sin \theta \cos \theta d\omega
d^2\rho =\int P_0(\vec \rho ,\omega )\sin \theta \cos \theta d\omega d^2\rho 
\end{equation}
expressing the energy conservation law: the radiation energy full flow
\-through the base plane{} remains constant regardless of scattering, as it
should be in case of real (dissipative) absorption absence. It is not
difficult to check that parity (\ref{eq24}) is valid for the exact solution
in the form (\ref{eq9}) and also for the solution in the form (\ref{eq21}).

{}With relation to the solution in the form (\ref{eq9}) the carried out
discussion discovers one curious peculiarity. It appears that the radiation
energy complete flow through the base plane is determined by the first term $%
\left( \stackrel{\sim }{P}\right) ${}. The second one $(X)$\ gives zero
contribution to the complete energy flow.

{}Let us investigate in more detail the structure of quantity $X(z,\vec \rho
,\omega )$\ {} in the invariant coordinate small-angle scattering
approximation. Proceeding to the Fourier-representation in the expression (%
\ref{eq13}) produces%
$$
\begin{array}{l}
X(z_0,\vec q,\omega )= 
\stackrel{z_0}{\stackunder{0}{\dint }}dz^{\prime }\int d\omega ^{\prime
}Q(z^{\prime };\omega ,\omega ^{\prime })\left[ \stackrel{\sim }{P}%
(z^{\prime },\vec q,\omega ^{\prime })-\stackrel{\sim }{P}(z^{\prime },\vec
q,\omega )\right] \cdot \\ \cdot \exp \left\{ i\vec q\left[ \vec \Phi
(z^{\prime };\omega ;\omega ^{\prime })+\vec D(z_0,0,\omega )\right]
\right\} 
\end{array}
$$
Employing regularity of function \ $\stackrel{\sim }{P}(z,\vec q,\omega )$,
the last expression can be written as%
$$
X(z_0,\vec q,\omega )=\stackrel{z_0}{\stackunder{0}{\dint }}dz^{\prime }%
\frac{\partial \stackrel{\sim }{P}(z^{\prime },\vec q,\omega )}{\partial
\omega }\stackrel{\sim }{A}(z^{\prime },\vec q,\omega )\exp \left[ i\vec
q\vec D(z_0,z^{\prime },\omega )\right] 
$$
where quantity \ {}$\stackrel{\sim }{A}(z,\vec q,\omega )$ is defined by (%
\ref{eq19}). Thus, it becomes evident that limit transition $\stackrel{\sim 
}{A}(z,\vec q,\omega )\longrightarrow 0$\ {} entails also $X(z_0,\vec \rho
,\omega )\longrightarrow 0$\ \ {}. This property has been established in
previous section with the aid of comparison of two variants of the REB
equation solution. Now we can see that its presence is determined by
structure of quantity\ $X(z_0,\vec \rho ,\omega )$.

{}Results of the present section give the weighty ground to believe that the
radiation spatial-and-angular distribution is determined basically by the
first term in the solution (\ref{eq9}). The second term represents the
amendment to the solution which can be neglected in the invariant coordinate
small-angle scattering approximation. This statement validity can be checked
under detailed research of properties of the obtained REB equation
approximated solutions by numerical methods.

\section{Technique of numerical calculation of multiple scattering effects}

We shall proceed from the obtained solution (\ref{eq15}) where only the
first term has been retained. In the considered approximation the multiple
scattering results in deformation of the radiation field reflected by a
plasma layer without change of a kind of function describing intensity
spatial and angular distribution. If only single ray with parameters $\theta
_0\left( \vec \rho \right) $, $\varphi _0\left( \vec \rho \right) $ comes
into each point $\vec \rho $ onto the base plane when reflecting from a
regular plasma layer (it will be so, if the source is dot-like and the
frequency is less than critical one for this layer), for function $P_0$ it
is possible to use expression of a kind 
\begin{equation}
\label{Eq8}P_0\left( \vec \rho ,\theta ,\varphi \right) =\stackrel{\sim }{P}%
_0\left( \vec \rho \right) \delta \left[ \cos \theta -\cos \theta _0\left(
\vec \rho \right) \right] \delta \left[ \varphi -\varphi _0\left( \vec \rho
\right) \right] 
\end{equation}
where quantity $\stackrel{\sim }{P}_0\left( \vec \rho \right) $ has the
meaning of energy flow at the point $\vec \rho $ in absence of scattering.
Substituting of (\ref{Eq8}) into (\ref{eq15}) and making integration over
angles one can obtain for the energy flow at the point $\vec \rho $ 
\begin{equation}
\label{Eq9}
\begin{array}{c}
\stackrel{\sim }{P}\left( \vec \rho \right) =\stackrel{\sim }{P}_0\left[
\vec \rho +\vec D\left( \theta _1,\varphi _1\right) \right] \cdot \\ \cdot
\mid \left\{ 1-\theta _{0\vec \rho }\left[ \vec \rho +\vec D\left( \theta
_1,\varphi _1\right) \right] \vec D_\theta \left( \theta _1,\varphi
_1\right) \right\} \cdot \\ 
\cdot \left\{ 1-\varphi _{0\vec \rho }\left[ \vec \rho +\vec D\left( \theta
_1,\varphi _1\right) \right] \vec D_\varphi \left( \theta _1,\varphi
_1\right) \right\} - \\ 
-\theta _{0\vec \rho }\left[ \vec \rho +\vec D\left( \theta _1,\varphi
_1\right) \right] \vec D_\varphi \left( \theta _1,\varphi _1\right) \varphi
_{0\vec \rho }\left[ \vec \rho +\vec D\left( \theta _1,\varphi _1\right)
\right] \vec D_\theta \left( \theta _1,\varphi _1\right) \mid ^{-1} 
\end{array}
\end{equation}
where $\vec D\left( \theta ,\varphi \right) \equiv \vec D\left( z_0,0;\omega
\right) $, subscripts $\vec \rho ,\theta ,\varphi $ mean derivatives with
corresponding variables, and $\theta _1,\varphi _1$ represent new arrival
angles of a ray.

Expression (\ref{Eq9}) uses explicit dependencies of arrival angles $\theta
_0,\varphi _0$ of a ray reflected from a regular plasma layer on position $%
\vec \rho $ which are usually unknown. As a rule, the dependence of
coordinates $\vec \rho $ on $\theta _0$ and $\varphi _0$ can be expressed in
an explicit form: $\vec \rho =\vec \rho _0\left( \theta _0,\varphi _0\right)
,$ where $\vec \rho _0\left( \theta ,\varphi \right) $ - point of arrival
onto the base plane of a ray with invariant angles $\theta $ and $\varphi $.
Expressing $\theta _{0\vec \rho }$ and $\varphi _{0\vec \rho }$ in (\ref{Eq9}%
) via $\dfrac{\partial \vec \rho }{\partial \theta }$ and $\dfrac{\partial
\vec \rho }{\partial \varphi }$, we obtain new representation for quantity $%
\stackrel{\sim }{P}\left( \vec \rho \right) $: 
\begin{equation}
\label{Eq10}\stackrel{\sim }{P}\left( \vec \rho \right) =\stackrel{\sim }{P}%
_0\left[ \vec \rho +\vec D\left( \theta _1,\varphi _1\right) \right] \left| 
\dfrac{\partial \left( \rho _{0x},\rho _{0y}\right) }{\partial \left( \theta
,\varphi \right) }\right| \left| \QDOVERD( ) {\partial \left( \rho
_{0x}-D_x,\rho _{0y}-D_y\right) }{\partial \left( \theta ,\varphi \right)
}\right| _{\theta =\theta _1,\varphi =\varphi _1}^{-1} 
\end{equation}

New arrival angles $\theta _1$ and $\varphi _1$ of a ray can be found by
solving the algebraic equation system 
\begin{equation}
\label{Eq11}\vec \rho -\vec \rho _0\left( \theta _1,\varphi _1\right) +\vec
D\left( \theta _1,\varphi _1\right) =0\text{ } 
\end{equation}

According to (\ref{Eq9}), an observer being at point $\vec \rho $ discovers
two effects connected with scattering in a plasma layer: change of the wave
arrival angles and change of the received signal intensity. The analytical
results of the present work are valid in a common case of magnetized plasma.
However, we shall consider below the case of an isotropic plasma linear
layer, as it displays the task basic features at relative simplicity of
numerical calculations. The simplification is mainly due to possibility to
define the ray trajectories in the analytical form. For a numerical
estimation of the effects described by the main term of the REB equation
solution it is necessary, first of all, to concretize the kind of the $%
\Delta N/N$ irregularity spectrum. We shall choose the spectrum of the
following kind: 
\begin{equation}
\label{Eq12}F\left( \vec \kappa \right) =C_A\left( 1+\kappa _{\perp
}^2/\kappa _{0\perp }^2\right) ^{-\nu /2}\delta \left( \kappa _{\Vert
}\right) 
\end{equation}
where $\kappa _{\bot }$ and $\kappa _{\Vert }$ are vector $\vec \kappa $
(irregularity spatial harmonic) components orthogonal and parallel
correspondingly to the magnetic field force lines, $\kappa _{0\perp }=2\pi
/l_{0\perp }$, $l_{0\perp }$ is the spectrum external scale, $\delta \left(
x\right) $ is the Dirac delta-function,%
$$
C_A=\delta _R^2\dfrac{\Gamma \left( \nu /2\right) }{2\pi \kappa _{0\perp }^2}%
\left[ \Gamma \QDOVERD( ) {\nu -2}{2}-2\QDOVERD( ) {R\kappa _{0\perp
}}{2}^{\left( \nu -2\right) /2}K_{\left( \nu -2\right) /2}\left( R\kappa
_{0\perp }\right) \right] ^{-1} 
$$
is a normalization constant, $\Gamma \left( x\right) $ is the
Gamma-function, $K_\beta \left( z\right) $ is the Macdonald function \cite
{SpecF}. The $\delta _R$ quantity characterizes the level of irregularities $%
\Delta N/N$. In mathematical theory of random fields it corresponds to the
structural function of the irregularity field for the scale length $R$ \cite
{Rytov78}. The considered model of a power-like small-scale irregularity
spectrum is used in many areas of modern physics. Both at the ionospheric F
region and at the tokamak plasma the irregularities are strongly stretched
along the magnetic field. We preserve this feature of irregularities even
when dealing with isotropic plasma model. Let the plasma layer with a linear
dependence of electron density on depth ($dz/dv=H$) be located at distance $%
h_0$ from the radiation source. The scattering cross-section for isotropic
plasma is \cite{Rytov78}: 
\begin{equation}
\label{Eq13}\sigma =\dfrac \pi 2k_0^4v^2F\left( \Delta \vec \kappa \right) 
\end{equation}
where $k_0=\omega /c$, $F\left( \vec \kappa \right) $ is the irregularity
spatial spectrum and $\Delta \vec \kappa $ is the scattering vector. For the
case of infinitely stretched irregularities the scattering vector
longitudinal and transversal components is defined by the expressions%
$$
\Delta \kappa _{\Vert }=k_0n\left( \cos \alpha -\cos \alpha ^{\prime
}\right) ; 
$$
$$
\Delta \kappa _{\bot }^2=k_0^2n^2\left[ \left( \sin \alpha \cos \beta -\sin
\alpha ^{\prime }\cos \beta ^{\prime }\right) ^2+\left( \sin \alpha \sin
\beta -\sin \alpha ^{\prime }\sin \beta ^{\prime }\right) ^2\right] 
$$
where notation of the first section is used and isotropic plasma refractive
index is $n=\sqrt{1-v}$. In a linear layer of isotropic plasma the vector $%
\vec \Phi $ components can be presented as%
$$
\Phi _x\left( v;\theta _k,\varphi _k,\theta _k^{\prime },\varphi _k^{\prime
}\right) =f\left( \stackrel{}{\theta _k^{\prime }}\right) \cos \varphi
_k^{\prime }-f\left( \theta _k^{}\right) \cos \varphi _k; 
$$
$$
\Phi _y\left( v;\theta _k,\varphi _k,\theta _k^{\prime },\varphi _k^{\prime
}\right) =f\left( \theta _k^{\prime }\right) \sin \varphi _k^{\prime
}-f\left( \theta _k^{}\right) \sin \varphi _k; 
$$
where%
$$
f\left( \theta _k\right) =2Hn\sin \theta _k\left( \sqrt{1-n^2\sin ^2\theta _k%
}+n\cos \theta _k\right) +h_0\sin \theta _k/\sqrt{1-n^2\sin ^2\theta _k}, 
$$
angles $\theta _k,\varphi _k,\theta _k^{\prime },\varphi _k^{\prime }$ are
current polar and azimuth angles of the wave vectors of incident and
scattered waves correspondingly at the ''vertical'' coordinate system.

Because of $\delta $-function presence in the irregularity spectrum (\ref
{Eq12}), the numerical estimation of expression for $\vec D\left( \theta
,\varphi \right) $ is reduced to calculation of double integral over $v$ and
one of angles. It is convenient to proceed to integration over angle $\beta
^{\prime }$ at the ''magnetic'' coordinate system. As result we obtain 
\begin{equation}
\label{Eq14}
\begin{array}{c}
D_x=\frac 12\pi k_0^3H^2C_A 
\stackrel{\cos ^2\theta }{\stackunder{0}{\dint }}dv^{.}v^2\sqrt{\dfrac{1-v}{%
\cos ^2\theta -v}}\stackrel{2\pi }{\stackunder{0}{\dint }}d\beta \{\{%
\underline{\sin \alpha _1\cos \beta }\cdot \\ \cdot \left[ 2\left( 
\sqrt{v+W_1^2\left( 1+v\right) }+W_1\sqrt{1-v}\right) +\dfrac{h_0}H\left(
v+W_1^2\left( 1-v\right) \right) ^{-1/2}\right] - \\ - 
\underline{\cos \varphi }\left[ 2\dfrac{\sin \theta }{\sqrt{1-v}}\left( \cos
\theta \pm \sqrt{\cos ^2\theta -v}\right) +\dfrac{h_0}H\dfrac{\sin \theta }{%
\sqrt{1-v}\cos \theta }\right] \}\cdot \\ \cdot \left[ 1+4k_0^2\kappa
_{0\bot }^{-2}\left( 1-v\right) \sin ^2\alpha _1\sin ^2\QDOVERD( ) {\beta
-\beta _1}{2}\right] ^{-\nu /2}+ \\ 
+\text{{\it similar term in which} }\alpha _1,\beta _1\text{{\it have been
replaced by} }\alpha _2,\beta _2\} 
\end{array}
\end{equation}
where $W_1=\cos \alpha _1\cos \gamma -\sin \alpha _1\sin \beta \sin \gamma $%
, angles $\alpha _1,\beta _1$ and $\alpha _2,\beta _2$ are polar and azimuth
angles of the incident wave wave vector at the ''magnetic'' coordinate
system for the trajectory ascending and descending branches correspondingly.
These are connected to the invariant angles by the relations%
$$
\cos \alpha _{1,2}=\pm \cos \gamma \sqrt{\dfrac{\cos ^2\theta -v}{1-v}}+%
\dfrac{\sin \theta }{\sqrt{1-v}}\sin \varphi \sin \gamma ; 
$$
$$
\sin \alpha _{1,2}=\sqrt{1-\cos ^2\alpha _{1,2}}; 
$$
$$
\sin \beta _{1,2}\sin \alpha _{1,2}=\dfrac{\sin \theta }{\sqrt{1-v}}\sin
\varphi \cos \gamma \pm \sqrt{\dfrac{\cos ^2\theta -v}{1-v}}\sin \gamma ; 
$$
$$
\cos \beta _{1,2}\sin \alpha _{1,2}=\dfrac{\sin \theta }{\sqrt{1-v}}\cos
\varphi 
$$
where the top signs correspond to the subscript 1, the bottom ones
correspond to the subscript 2. Expression for $D_y$ is derived from the
expression for $D_x$ by replacement of underlined factors \underline{$\sin
\alpha _{1,2}\cos \beta $} with $\left( \cos \alpha _{1,2}\sin \gamma +\sin
\alpha _{1,2}\sin \beta \cos \gamma \right) $ and \underline{$\cos \varphi $}
with $\sin \varphi $ correspondingly. Two terms in (\ref{Eq14}) correspond
to the trajectory ascending and descending branches of a ray with
coordinates $\theta ,\varphi $. The equation system (\ref{Eq11}) numerical
solving was performed using the Newton's globally converging method
described, for example, in \cite{Dennis88}.

\section{Calculation results for isotropic plasma}

The calculations were carried out for the following set of parameters: $h_0$
= 150 km, $H$ = 100 km, $\nu $ = 2.5, $l_{0\bot }$ = 10 km, $R$ = 1 km, $%
\delta _R=3\cdot 10^{-3}$ (ionospheric irregularity level characteristic of
night quiet conditions), frequency $f$ = 5 MHz, the angle of irregularity
inclination (''magnetic field'' inclination) $\gamma =25^o$. Thus we mean
the conditions of the ionosphere sounding from the Earth surface. The
intensity attenuation and change of the arrival angles of a signal reflected
from a layer (with relation to those values when reflecting from the same
plasma layer without irregularities) were calculated for area which sizes
were 800 km along the $y$ axis (the magnetic meridian direction) and 400 km
along the $x$ axis. The radiation source was in the coordinate origin. The
calculation results are represented in figures 2 - 4. On the contour map of
fig. 2 the constant level lines of the received signal intensity attenuation
(in dB) are shown. The signal attenuation calculated value is symmetric with
relation to both the $y$ axis (i.e. magnetic meridian plane) and the $x$
axis with accuracy determined by numerical calculation errors. The result is
not trivial: there is asymmetry in the problem conditions due to the
irregularity inclination. The central symmetry is required by the
reciprocity theorem \cite{Ginz67,Bud61}. Thus the obtained numerical
solution is in accordance with the electromagnetic field general properties.
This is an additional argument to the benefit of the conclusion about the
primary significance of the first term in the REB equation approximate
solution (\ref{eq15}).

The main detail in fig. 2 is the region having the shape of ellipse with
half-axes of 300 km and 60 km where significant intensity attenuation (up to
15 dB) takes place. Outside of this region some increase of the signal
intensity (in comparison with its value in absence of irregularities) is
observed. This is quite natural result because at an nonabsorbing medium the
complete radiation energy flow is conserved and scattering results only in
its spatial redistribution. At larger distance from the source the intensity
change aspires to zero. The transition from the region of the reflected
signal attenuation to the region of amplification is of sharp character.
That is, probably, a consequence of approximation used under transformation
of the expression (\ref{eq14}) to the expression (\ref{eq15}) fist term. One
can expect that retaining of the higher order terms in the series expansion
will result in smoothing of the above transition.

The second of multiple scattering effects (i.e. change of the arrival
angles) is illustrated if figs. 3 and 4. In fig. 3 the contour lines show
the absolute value of the polar arrival angle $\theta $ alteration. In fig.
3 the alteration of azimuth angle is presented. One can see that distortion
of the polar angle reaches of $5$ and maximum alteration of the azimuth
angle is $90^o$.

Both effects are observed in experiment. The intensity reduction of the
vertical sounding signal reflected from the ionosphere should be interpreted
by the observer as an additional collisionless mechanism of radio wave
attenuation. This phenomenon is observed under natural conditions bringing
the increased values of the effective collision frequency \cite
{Setty71,Denis83}. There are the weighty grounds to believe that it is
connected to development of small-scale irregularities in the ionosphere. In
particular, it is displayed stronger at night time and can reach of 10-15 dB 
\cite{Denis83}. The latter figure is in accordance with our calculation
results. The experimental data on the arrival angle change of a wave on a
short line (when transmitter to receiver distance is about 100 km) are
reported in \cite{Baulch84}.

\section{Conclusion}

In the present work the heuristic basis for use of the invariant coordinate
small-angle scattering approximation under solving of the RTE for a
magnetized plasma layer is considered. Within the framework of this
approximation two versions of the analytical solution have been obtained.
They describe spatial-and-angular distribution of radiation reflected from a
monotonous plasma layer with small-scale irregularities.

The physical conclusions about influence of the multiple scattering effects
in a layer of plasma on the spatial-and-angular characteristics of radiation
are possible on the basis of detailed numerical research of the obtained
solutions. Such research has been carried out in the present work for the
case of isotropic plasma.

It was shown that the main term of the REB equation solution for the
radiation reflected from a plasma layer with random irregularities describes
two effects: the signal intensity change (attenuation for the normal
sounding) and the arrival angle change. Both effects are observed in the
experiments on the ionosphere radio sounding. The first one is known as
anomalous attenuation of radio waves at natural conditions. Note, that under
the ionosphere heating experiments another kind of anomalous attenuation is
observed: its mechanism is based on the mode transformation under
scattering, not on the multiple scattering \cite{Robin89}. The effect of
arrival angle change can be interpreted as a mechanism of additional
refraction in the ionosphere and also has experimental confirmation. These
two effects numerical estimations obtained in the present work for
parameters of a plasma layer and irregularity spectrum typical for
mid-latitude ionosphere are in accordance with experimental data.

The considered effects can be observed not only at the ionosphere radio
sounding but also at sounding by electromagnetic radiation of other kinds of
plasma with random irregularities both in natural and in laboratory
conditions.

{\em Acknowledgments.} The work was carried out under support of Russian
Foundation of Basic Research (grants No. 94-02-03337 and No. 96-02-18499).

\begin{center}
{\sf \newpage }Figure captions
\end{center}

Fig. 1. A schematic plot of Poeverlein's construction for ordinary waves.
The refractive index surfaces for several values of $v$ ($v=\omega
_p^2/\omega ^2$) are shown. The ray trajectories in this ''$k$-space'' are
represented by straight dashed lines parallel to $z$ axis.

Fig. 2. Contour map showing the relative alteration of the intensity (in dB)
of the signal reflected from the ionospheric plasma layer due to multiple
scattering. Radiation source is in the coordinate origin.

Fig. 3. Contour map showing the relative alteration of the polar arrival
angle (in degrees) of the signal reflected from the ionospheric plasma layer
due to multiple scattering.

Fig. 4. Same as in fig. 3, but for the azimuth arrival angle.

\end{document}